\documentclass[journal]{IEEEtran}
\usepackage{amsmath,amssymb,amsfonts}
\usepackage{graphicx}
\usepackage{multirow}
\usepackage{makecell}

\ifCLASSINFOpdf

\else

\fi

\begin{document}

\title{Flexible Piezoresistive Yarn Sensor for Human Physiological Signal Measurement}

\author{Rizal Maulana,
        Ádám Rák,
        Sándor Földi,
        and György Cserey,~\IEEEmembership{Member, IEEE}
\thanks{R. Maulana, Á. Rák, S. Földi, and G. Cserey are with Faculty of Information Technology and Bionics, Pázmány Péter Catholic University, Budapest, Hungary}}

\maketitle

\begin{abstract}
Continuous monitoring of physiological signals is essential for the early detection of health problems. A measurement system that ensures high sensitivity, accuracy, and user comfort is needed. In this study, we designed and optimized a flexible piezoresistive yarn (FPY) sensor to achieve a high sensitivity and wide working range for detecting physiological signals. The representative sensor design was constructed by applying an FPY bonding pattern, utilizing tightly arranged triangular patterns and using minimal FPY. The prototype sensor operates in two measurement modes, strain and pressure, and was evaluated for measuring neck motion, finger bending, respiratory signals, and arterial blood pressure (ABP) waveforms. A qualitative evaluation, performed by comparing the characteristics of the measurement results of each physiological signal with those from related studies, indicates a high similarity in its morphological characteristics. Then, a quantitative evaluation through baseline drift analysis demonstrates that the FPY sensor displays high measurement stability. The ABP waveform measurement shows the most stable baseline, with a mean absolute error (MAE) of 0.0051 ± 0.0029 in terms of baseline drift, using normalized values from 0 to 1. Based on our results, the prototype sensor can be used as an innovative solution for physiological signal monitoring and can be further enhanced for personalized healthcare and sports applications.
\end{abstract}

\begin{IEEEkeywords}
Continuous monitoring, flexible piezoresistive yarn, physiological signal, sensor design optimization.
\end{IEEEkeywords}

\IEEEpeerreviewmaketitle

\section{Introduction}
\label{sec:introduction}
\IEEEPARstart{F}{lexible} sensors, renowned for their flexibility, adaptability, and conformability, are at the forefront of sensor technology development. The characteristics of flexible sensors shift the typical perception of sensing, which is commonly associated with the rigid sensor. Flexible sensors enhance comfort and allow precise sensor placement, as they can conform to contours while providing close contact with the measurement point. Due to these advantages, flexible sensors have been widely implemented in various applications, including human-machine interfaces \cite{ref-1,ref-2,ref-3}, bioelectronics \cite{ref-4,ref-5,ref-6}, and healthcare \cite{ref-7,ref-8,ref-9}. The implementation of flexible sensors as wearable devices in healthcare facilitates continuous monitoring of physiological signals within the human body.

In healthcare applications, continuous monitoring of physiological signals is essential. This monitoring provides an in-depth understanding of the various physiological cycles in the body and their direct correlation as indicators of human health \cite{ref-10}. Sudden and unusual changes in some physiological signals, such as respiratory rate and blood pressure, potentially become dangerous indicators for human health. Continuous monitoring can facilitate the early detection of these abnormalities. Early detection plays a vital role in providing timely and appropriate treatment to the sufferer, potentially minimizing the risks resulting from these conditions.

Earlier than the development of recent sensor technology, the measurement of physiological signals relied on conventional methods. While capable of providing highly accurate measurement results, this conventional method possesses several limitations. The measurement process requires professional medical assistance, and the measurement positions and conditions are constrained due to the commonly bulky dimensions of the instruments \cite{ref-11}. These characteristics make the conventional method non-wearable, limiting the application in continuous monitoring of physiological signals.

The current development of flexible sensor technology represents a significant step in developing a system for continuous monitoring of physiological signals. These developments are inseparable from the advancements in the material sciences. The optimization of material properties paves the way for the fabrication of sensors with compact dimensions and lightweight design while maintaining high flexibility and sensitivity. These features enable sensors to be applied as a wearable device with a high level of accuracy.

In wearable healthcare applications, strain and pressure sensors are the most commonly used for monitoring physiological signals. These sensors are employed to detect mechanical changes in the body resulting from the physiological cycle \cite{ref-12,ref-13}. Strain and pressure sensors are classified into four categories based on their transduction mechanism: piezoelectric \cite{ref-14,ref-15}, triboelectric \cite{ref-16,ref-17}, capacitive \cite{ref-18,ref-19}, and piezoresistive \cite{ref-20,ref-21}. Piezoresistive, which operates by changing their electrical resistance in response to structural deformation caused by applied mechanical stress, provides numerous advantages. They have greater adaptability to fluctuations in environmental conditions, a straightforward fabrication process due to their simple structure, and are also characterized by low-power consumption \cite{ref-22,ref-23}.

Current technology enables the fabrication of flexible piezoresistive sensors in various forms, including flexible sheets \cite{ref-24,ref-25}, fabrics \cite{ref-26,ref-27}, and yarns \cite{ref-28,ref-29}. Among these, flexible sheets have the lowest level of flexibility due to their layered structural design. Additionally, flexible sheets are limited to working only as pressure sensors. In comparison, flexible fabric and yarn has similar flexibility at a high level. Nevertheless, flexible yarn offers the additional advantage of applying to a broader range of mediums. These sensors can be seamlessly integrated into daily wear through sewing or embroidery. Flexible yarn can also be used directly without any additional base material by configuring it into specific bond patterns, which can be optimized according to measurement requirements. In addition, flexible yarn enables operating both as strain and pressure sensor.

Numerous implementations of flexible piezoresistive yarn (FPY) sensors for multiphysiological signal monitoring have been performed, involving those applying cytoskeleton-inspired core-shell yarn (CCSY) \cite{ref-30} and graft-modified multi-walled carbon nanotubes (g-MWCNTs) \cite{ref-31}. The fabrication process for both CCSY and g-MWCNTs is challenging, as it requires technical adjustment, resulting in more extensive and complex procedures. In addition, both designed sensors have the same limitation concerning their working range. Both were evaluated based on physiological signal measurements characterized by relatively high frequencies, including pulse waves and joint movements. The evaluation of low-frequency physiological signals, such as respiratory signals, is missing. The working range of the sensor can be enhanced through design optimization. In the CCSY implementation, flexible yarn is integrated into the fabric by embroidery, forming a plus sign pattern. Meanwhile, flexible yarn is used simply as a single strand in the application of g-MWCNTs. Both designs are suboptimal in terms of sensor placement accuracy and sensor coverage area at the designated measurement point. These attributes influence both their working range and sensitivity. Therefore, a proper sensor design is crucial to achieve a wide working range and enhanced sensor sensitivity.

In this study, we developed a FPY sensor based on conductive thermoplastic polyurethane (TPU) filament. A simple extrusion technique was used to process the filament material. The sensor design optimization was performed to meet the optimal sensor characteristics. The designed prototype sensor has the capability to operate both as a strain and pressure sensor. This sensor was evaluated for measuring diverse physiological signals at various locations in the human body, including motion related to different events in the neck, finger bending, respiratory signals, and arterial blood pressure (ABP) waveform.

The remaining section of this manuscript is organized as follows. In Section \ref{sec:methods}, the materials and methods used are introduced, including the preparation, preliminary pattern design, and final design of the FPY sensor, as well as the measurement protocol used to validate its capabilities for physiological signal measurement. In Section \ref{sec:results}, the results of sensor pattern design optimization and measurement of each physiological signal, along with qualitative and quantitative analysis, are presented, with conclusions drawn in Section \ref{sec:conclusions}.

\section{Materials and Methods}
\label{sec:methods}
\subsection{Preparation of the Flexible Piezoresistive Yarn (FPY)}
The base material for the fabrication of FPY is conductive TPU filament (Recreus, Alicante, Spain) containing conductive graphene particles. The fabrication process utilized a 3D printer to perform a simple extrusion technique. The initial procedure required is to dry the filament to achieve extrusion results without bubble formation. Additionally, a high extrusion speed is also necessary to prevent this condition. In the extrusion process, the temperature is maximized without burning the filament to produce more consistent FPY resistance. The extrusion process employs the hanging loop method, creating a loop in the extruded filament while maintaining a consistent distance between the loop and the 3D print nozzle.

\subsection{Preliminary Pattern Design for Flexible Piezoresistive Yarn (FPY) Sensor Sensitivity Optimization}
In an FPY-based sensor, the sensor pattern design significantly affects its sensitivity. This study analyzes this effect by focusing on two main parameters of sensor pattern design: pattern density and FPY length. Both parameters are investigated by varying them during the implementation of a representative triangular stitch pattern for FPY in elastic fabric. This preliminary pattern design aims to evaluate sensor sensitivity across the full range of the applicable forces, ensuring that the elastic fabric can be stretched to its maximum without compromising the stitching pattern.

\begin{figure*}[t]
    \centering
    \includegraphics[width=5.8in]{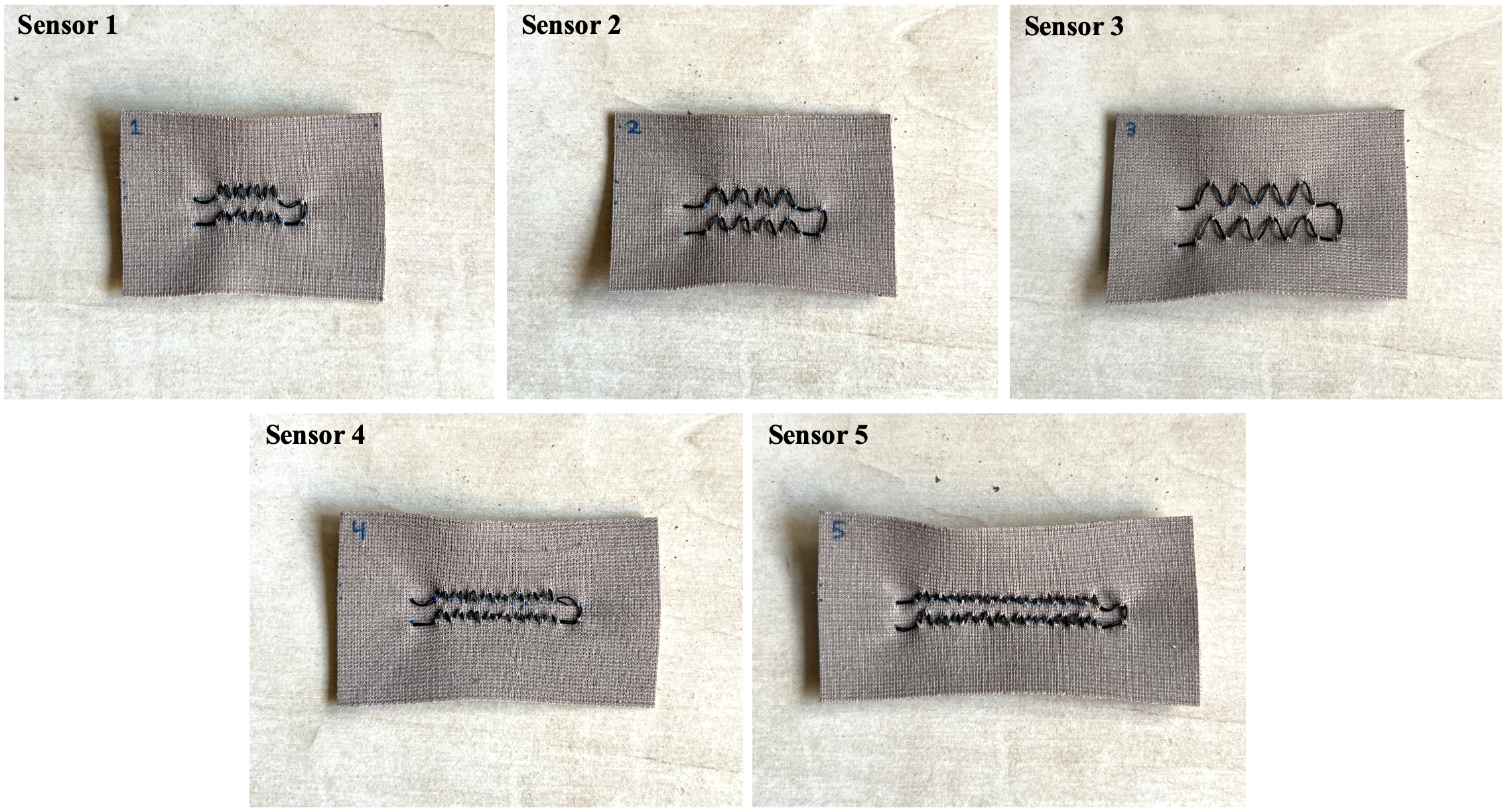}
    \caption{Sensor pattern design configuration involves varying the pattern width and FPY length, implemented as a triangular stitch pattern on elastic fabric.}
    \label{fig1}
\end{figure*}

The sensor pattern design is implemented by stitching two rows of a triangular pattern onto elastic fabric. The pattern density varied by modifying the widths of the triangular pattern, while the FPY length was adjusted by varying the number of triangular patterns. Five sensor configurations were designed, as shown in Fig. \ref{fig1}. Sensor 1 features the tightest design and uses the shortest FPY, with a triangular pattern width of 2 mm and four triangular patterns in each row. Sensor 2 and sensor 3 have a looser configuration, each including four triangular patterns in each row, with triangular pattern widths of 4 mm and 6 mm, respectively. Sensor 4 and sensor 5 have a longer FPY used, designed with a triangular pattern width of 2 mm, having eight and twelve triangular patterns in each row, respectively.

Sensor pattern design evaluation was performed on two mechanisms: strain testing and pressure testing. Strain testing was conducted by stretching the sensor by four different lengths: 0.5 cm, 1.0 cm, 1.5 cm, and 2.0 cm. The pressure testing involved manually applying pressure to the center of the sensor at three different levels: low, medium, and high. In both tests, each increase in the applied force was preceded by returning the sensor to its initial condition.

\subsection{Final Design of the Flexible Piezoresistive Yarn (FPY) Sensor}
The final design of the FPY sensor was implemented by configuring FPY into a bonding pattern. Applying parameters obtained from the sensor pattern design optimization results presented in Section \ref{sec:results}, along with design adjustment concerns.

From a design standpoint, the establishment of this bonding pattern enhances stability in sensor placement and sensor coverage area at the measurement point, especially compared to utilizing an FPY as a single strand without pattern \cite{ref-31}. In addition, the application of bonding patterns also enhances the physical robustness of the sensor and increases its resistance to noise from small motion disturbances. As a result, this configuration could ensure that changes in the sensor’s electrical resistance are only influenced by the intended applied strain or pressure.

\begin{figure*}[t]
    \centering
    \includegraphics[width=5.8in]{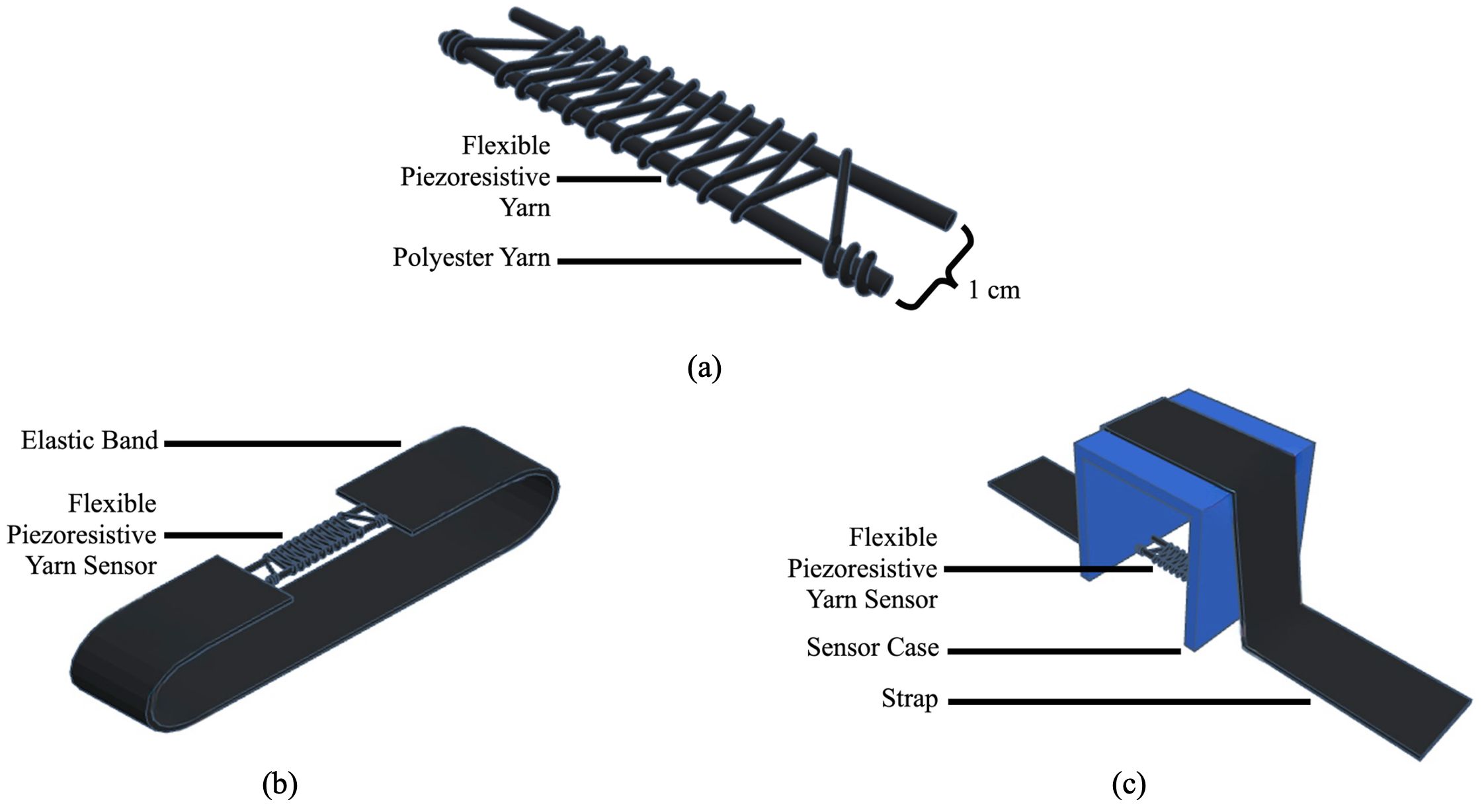}
    \caption{Schematic figure of the sensor design. (a) FPY sensor design with a tight triangular bonding pattern. (b) Strain mode design with an additional elastic band. (c) Pressure mode design with sensor case and strap.}
    \label{fig2}
\end{figure*}

The assembly of the FPY bonding pattern was conducted using two parallel polyester yarns. Both polyester yarns are represented as 3 cm long and 1 cm apart. The selection of polyester yarn is based on its flexibility and its insulating properties, so it does not affect the electrical resistance of the sensor. Based on our experiments, the optimal sensitivity is achieved by utilizing a tight pattern and a minimal length of FPY. An additional consideration is the area of the measurement point. By considering all these parameters, we developed a sensor design featuring a tight triangular bonding pattern with a representative of eleven triangles, as shown in Fig. \ref{fig2}a. The design enables the sensor to operate in two measurement modes: strain and pressure.

\subsubsection{Strain Mode Design}
The sensor design for the strain mode only requires two additional elastic bands connected through loop fasteners. A designed FPY sensor is attached at each end to an elastic band, as shown in Fig. \ref{fig2}b. The use of elastic bands aims to facilitate physiological signal measurements over a wider area, thereby minimizing the need for longer FPY, which can reduce sensor sensitivity. Additionally, employing loop fasteners enables the adjustment of the elastic band by the width of the body part where the measurement point is located.

The strain mode works on the principle of the piezoresistive effect in FPY. This effect occurs when mechanical stress is applied to the FPY, causing alterations in its conductive particle network structure and resulting in changes in its electrical resistance. The electrical resistance of FPY increases with tension and decreases under compression. The impact of the piezoresistive effect is additionally affected by the integration of conductive particles within the TPU matrix.

This strain mode design configuration can be applied to measure physiological signals that generate physical motion in the body. The physical motion occurring within the measurement area may cause stretching and deformation of the FPY sensor and change its electrical resistance. In this study, we measure several physiological signals using a strain mode sensor, including motion related to different events in the neck, such as coughing, swallowing, and talking; finger bending; and respiratory signals in the abdomen.

\subsubsection{Pressure Mode Design}
In the pressure mode design, two additional components are used: a sensor case and a strap. The prototype of the sensor case is designed in the form of hollow block objects with its front, back and bottom made open. The FPY sensor is mounted to the bottom of the sensor case to ensure direct contact with the pressure source. On the outer part of the sensor case, an adjustable strap is added, which serves to facilitate sensor use and maintain the sensor position stability at the measurement point. Fig. \ref{fig2}c shows the design of the pressure mode sensor.

The pressure mode also works based on the piezoresistive effect in FPY. In this mode, resistance change relates to the movement of conductive particles within the TPU matrix due to contact on the FPY surface. Under typical conditions, the particles form a conductive network with a baseline electrical resistance. The electrical resistance increases when the FPY is subjected to pressure, which causes the conductive particles to move apart and vice versa. The magnitude of the electrical resistance change is influenced by the concentration and distribution of conductive particles within the TPU matrix.

In this study, the pressure mode is used to measure the ABP waveform at the arterial site on the wrist. ABP waveforms are characterized by low-magnitude signals, requiring a specific approach for the detection. In our design, we implemented an applanation method to amplify the arterial pressure, making it detectable by the FPY sensor. The applanation method involves applying light pressure on the arterial site to flatten the arterial surfaces \cite{ref-32}. The vertical force generated at that site is proportional to the pressure on the artery.

The applanation process in the designed sensor involves the interaction of its components, including the polyester yarn, sensor case, and strap. The flexibility of polyester yarn allows the FPY sensor to conform to the wrist’s curvature. The tightening of the sensor case by strap enables the polyester yarn to apply light pressure on the arterial sites on both sides. The area between the two parallel polyester yarns, designated for constructing the FPY bonding pattern, is the optimal area for detecting amplified arterial pressure.

\subsection{Measurements and Characterizations}
\subsubsection{Measurement Protocols}
The data collection for various physiological signal measurements involved a single healthy male subject aged 35 years. The subject was instructed to remain seated and avoid excessive movements during the measurement process. The FPY sensor was placed on the body part corresponding to the physiological signal being measured. The output of the FPY sensor, which is in the form of electrical resistance, was converted into a voltage output using a voltage divider circuit with a pull-up resistor of 1 M{$\Omega$}. An oscilloscope (Keysight, California, USA) is used to record the measurement data. In its application as a wearable device, the oscilloscope can be replaced by a microcontroller module that additionally performs data processing. The sampling frequency of the measurements was 500 Hz.

Multiple filters were used in signal processing, including notch filters, low-pass filters, and moving average filters. The notch filter was implemented to eliminate 50 Hz noise caused by power line interference. A low-pass filter with a different cut-off frequency, 10 Hz for neck motion and finger bending measurements or 15 Hz for respiratory signal and ABP waveform measurements, was used to remove high-frequency noise. Then, a moving average filter was employed to reduce motion artifacts in the measurement signals. The aforementioned procedures were performed using MathWorks MATLAB 2024b. The measurement results of each physiological signal were validated by comparing its characteristics with those of related signals from other studies. In addition, baseline drift and its percentage relative to amplitude for each measurement were calculated to evaluate sensor performance.

\subsubsection{Neck Motion}
\begin{figure}[t]
    \centering
    \includegraphics[width=3.5in]{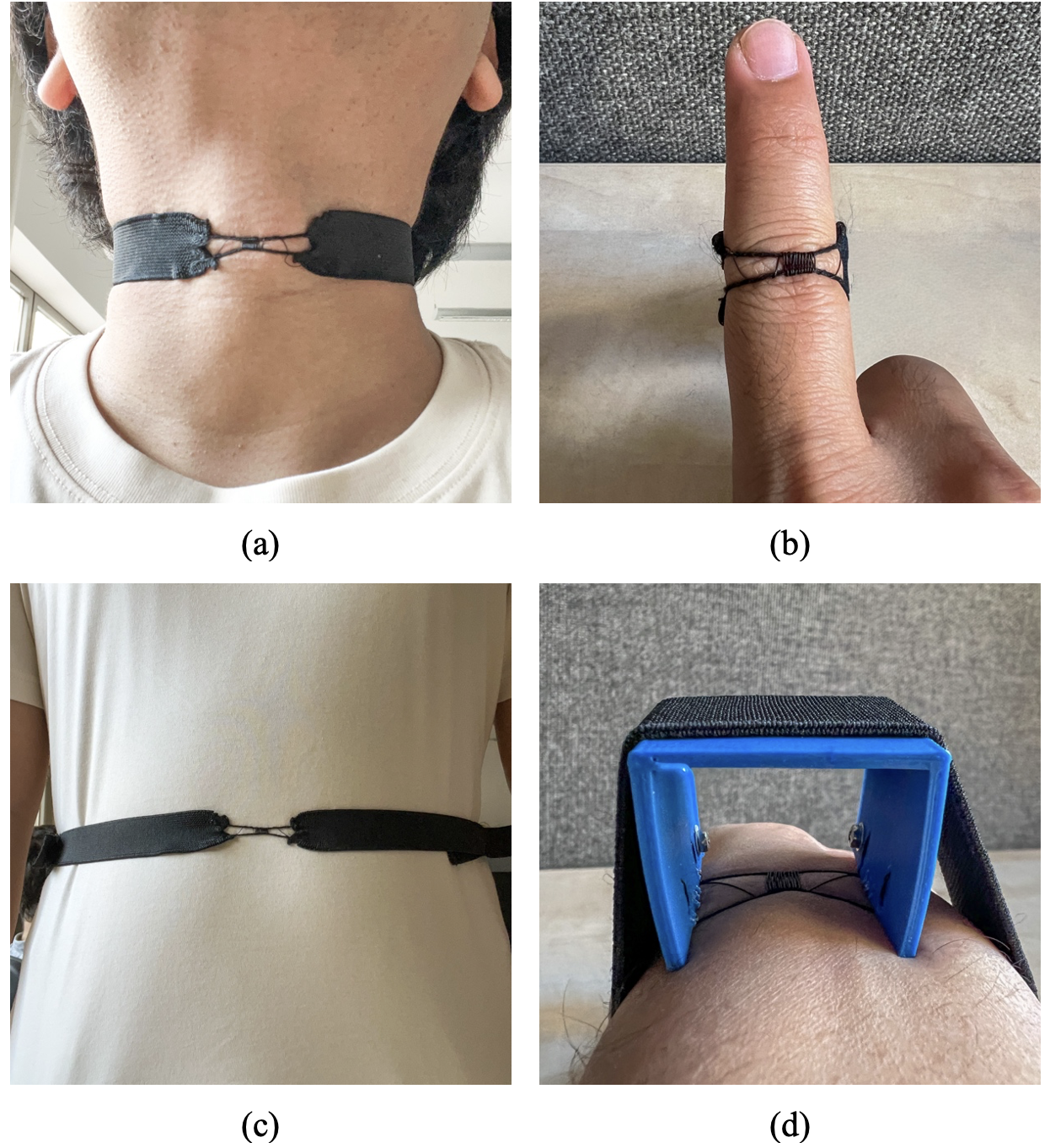}
    \caption{FPY sensor placement for each physiological signal measurement. (a) Neck motion. (b) Finger bending. (c) Respiratory signal. (d) ABP waveform.}
    \label{fig3}
\end{figure}

The measurement of neck motion aims to detect physiological motions resulting from throat vibrations during coughing, swallowing, and talking events. In this measurement, a strain mode FPY sensor is used as a choker around the neck, as shown in Fig. \ref{fig3}a. The FPY bonding pattern part is positioned at the location of the Adam’s apple, which moves up and down during throat vibrations, causing deformation of the sensor.

In the experiment on cough detection, the subject was instructed to intentionally cough five times with consistent intensity in a single measurement cycle. Previous studies indicate that cough signals from strain sensors have a single dominant peak with a nonreproducible waveform due to variations in neck muscular activity that are not identical in each coughing event \cite{ref-33,ref-34}. Then, on swallowing detection, a single swallowing event was performed during a single measurement cycle. The swallowing signal is characterized by two peaks within a single signal period, with the second peak having a greater amplitude. In normal conditions, the period of the swallowing signal ranges from 1.5 to 2 seconds \cite{ref-35,ref-36}. As for the talking detection, three different words were pronounced in the experiment: ``strain'', ``sen-sor'', and ``mo-ni-tor''. Each word is pronounced once in a single measurement cycle. Talking signal is defined by the number of signal peaks corresponding to the number of syllables pronounced \cite{ref-37,ref-38}.

\subsubsection{Finger Bending}
Finger bending measurements are intended to detect changes in finger joint angles. The FPY strain sensor is used to detect these changes by being worn as a ring on the index finger positioned on the knuckle, as shown in Fig. \ref{fig3}b. The experiment involved examining three different finger joint angles: 30°, 60°, and 90°. The experiment was performed by applying the same finger joint angles five times during a single measurement cycle. The finger bending signal is characterized by an amplitude that increases proportionally with the angle of the finger joints \cite{ref-22,ref-30,ref-37,ref-38,ref-39}.

\subsubsection{Respiratory Signal}
Respiratory signal measurement aims to continuously and non-invasively monitor respiratory activity based on body movements during breathing. In this study, respiratory signals were measured at the abdomen. The FPY strain sensor was applied to the abdomen circumference with the FPY bonding pattern part centered at that site, as shown in Fig. \ref{fig3}c. The expanding and contracting movements at the abdomen during inhalation and exhalation cause the FPY sensor to stretch. To achieve optimal measurement results, the attachment of loose elastic bands should be avoided, as it reduces the stretching effect on the FPY sensor.

In the experiment, we evaluated three types of breathing: normal, deep, and rapid. A single measurement cycle was conducted over a 10-second period, during which normal, deep, and rapid breathing was performed three, two, and six times, respectively. The characteristics of the respiratory signal, specifically its frequency and amplitude, depend on the type of breathing. Deeper breathing results in greater amplitude and lower frequency, and vice versa \cite{ref-40,ref-41}.

\subsubsection{Arterial Blood Pressure (ABP) Waveform}
The ABP waveform measurements are intended to facilitate continuous non-invasive blood pressure waveform monitoring by sensing arterial pressure at the wrist. The FPY pressure mode sensor is worn on the wrist, and the FPY bonding pattern part must be precisely placed on the arterial site, as shown in Fig. \ref{fig3}d. High-precision sensor placement is necessary for the sensor to accurately detect arterial pressure at a specific measurement point. Additionally, proper strapping of the sensor is also crucial to ensure consistent applanation pressure during measurement, thereby enabling the sensor to provide a stable output signal. In our experiments, each measurement was performed for 10 seconds with the subject’s hand lying on the table. In normal conditions, the ABP waveform has 60-100 signal cycles per minute, corresponding to the number of normal heartbeats. According to its signal morphology, the ABP waveform is characterized by the presence of a systolic peak, a dicrotic notch, and a dicrotic peak within a single signal period \cite{ref-32,ref-42}.

\subsection{Statistical Methods}
The baseline stability of the measured signal was assessed by evaluating the baseline drift from each measurement. These evaluations were performed by calculating the difference between the baseline values of each signal cycle and the initial baseline values from a single measurement. The baseline value used in the calculation is the measured voltage value that has been normalized with a range of 0 to 1, with reference to all measurement results. Each type of physiological signal measurement used data from five repeated measurements. The average baseline drift calculation used Mean Absolute Error (MAE), as defined in \eqref{eq1}.

\begin{equation}
    MAE=\frac{1}{n} \sum_{i=1}^{n}\left|y_{i}-x_{i}\right|,\label{eq1}
\end{equation}
where {$n$} is the number of signal cycles in a single measurement, {$y_i$} is the initial baseline value, and {$x_i$} is the baseline value of the {$n$}-th cycle. In addition, a standard deviation (SD) was computed to assess variability.

The average and SD of the signal amplitude value were calculated to evaluate the proportionality between the magnitude of mechanical movement of the body and the signal detected by the sensor. The signal amplitude value was derived from the difference between the peak value of a particular cycle and its baseline value. Last, the average percentage of baseline drift relative to the amplitude for each signal cycle was computed. The computation of relative baseline drift (RBD) is described in \eqref{eq2}.

\begin{equation}
    RBD=\left(\frac{b_{n}-b_{0}}{a_{n}}\right) \times 100 \%,\label{eq2}
\end{equation}
where {$b_n$} is the baseline value after the {$n$}-th cycle, {$b_0$} is the initial baseline value, and {$a_n$} is the signal amplitude value of the {$n$}-th cycle.

\section{Results and Discussion}
\label{sec:results}
\subsection{Sensor Pattern Design Optimization}
\begin{figure*}[t]
    \centering
    \includegraphics[width=4.7in]{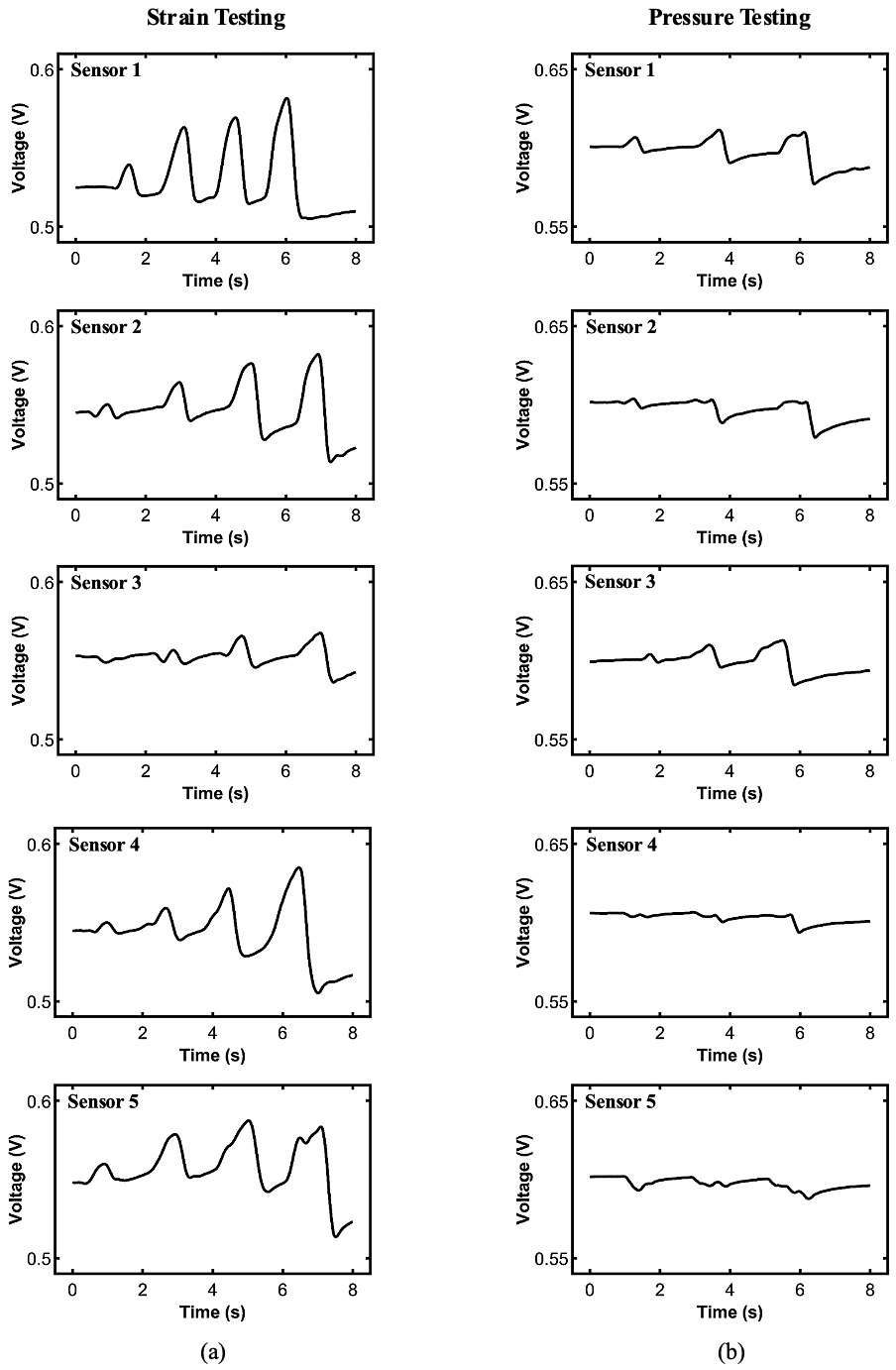}
    \caption{Measurement results of sensor pattern design evaluation. (a) Strain testing with four different stretching levels. (b) Pressure testing with three different pressure levels.}
    \label{fig4}
\end{figure*}

The strain testing results, used to evaluate the piezoresistive effect underlying the strain mode principle, are shown in Fig. \ref{fig4}a. Sensor 1 shows the greatest sensitivity, as shown by the greatest signal amplitude compared to the other sensors at each strain level. In terms of signal morphology, sensor 1 displays the most distinctly recognizable amplitude increase, which is directly proportional to an increase in strain level. Additionally, the signal output from sensor 1 is the most accurate for returning to the baseline once it returns to its initial condition without strain. In general, as the sensor design becomes looser, the sensing ability to detect small strains decreases. This condition is seen in sensor 3, which was unable to detect a strain of 0.5 cm. The decrease in sensing ability is also observed in sensors using longer FPY, although the impact is less significant than a looser design sensor.

The pressure testing result, which correlates with the piezoresistive effect in pressure mode, shown in Fig. \ref{fig4}b, indicates that sensor 1 has slightly greater sensitivity compared to the sensors with a looser design. This small sensitivity gap occurs because all sensors with varying pattern density have an area affected by pressure and the total sensor area is not significantly different. The difference in triangular pattern width among sensor 1, sensor 2, and sensor 3 is only 4 mm. However, a tighter sensor design offers further advantages in its implementation in pressure measurement at narrow measurement points. These sensors have greater precision in the contact between the sensor pattern and the pressure source. The difference in signal output is more distinctly observed on the sensors with varying FPY length. The longer the FPY used, the smaller the ratio of the sensor area affected by pressure to the total sensor area, resulting in a decrease in the sensitivity of the sensor in detecting the pressure.

Sensor 1, identified as the most optimal sensor design under the tested conditions, was evaluated in physiological signal measurements. However, the measurement results signal obtained was not optimal, especially in a low-magnitude pressure signal. The utilization of elastic fabrics causes an attenuation in the pressure measured by the FPY sensor. Based on the overall experiment results, the sensor design was optimized as an FPY bonding pattern on polyester yarn to eliminate the attenuation effect on elastic fabrics. The bonding pattern design remains focused on implementing the tightest pattern and shortest FPY used. The higher flexibility in forming a bonding pattern than a stitching pattern enables the implementation of a tighter sensor design. However, there is a limitation preventing the triangular patterns from being designed so tightly that they stick to each other. This condition could cause a significant change in the electrical resistance of the FPY sensor with a small force applied. This should be avoided to minimize measurement noise from small forces that are not generated from the main strain or pressure source.

The establishment of the minimum FPY length used, specified by the number of triangular patterns, needs to consider the coverage of the measurement point area \cite{ref-43}. The pressure testing results analysis determined that sensor sensitivity is greatly affected by the ratio of the sensor area affected by pressure to the total sensor area. Optimal sensitivity can be obtained by maximizing this ratio, provided that the sensor pattern dimensions adequately cover the measurement point area. Considering each measurement point in the physiological signal measurements performed, the final sensor design was implemented as a representative of a single row of eleven triangular bonding patterns.

\subsection{Neck Motion Measurement}
\begin{figure*}[t]
    \centering
    \includegraphics[width=7in]{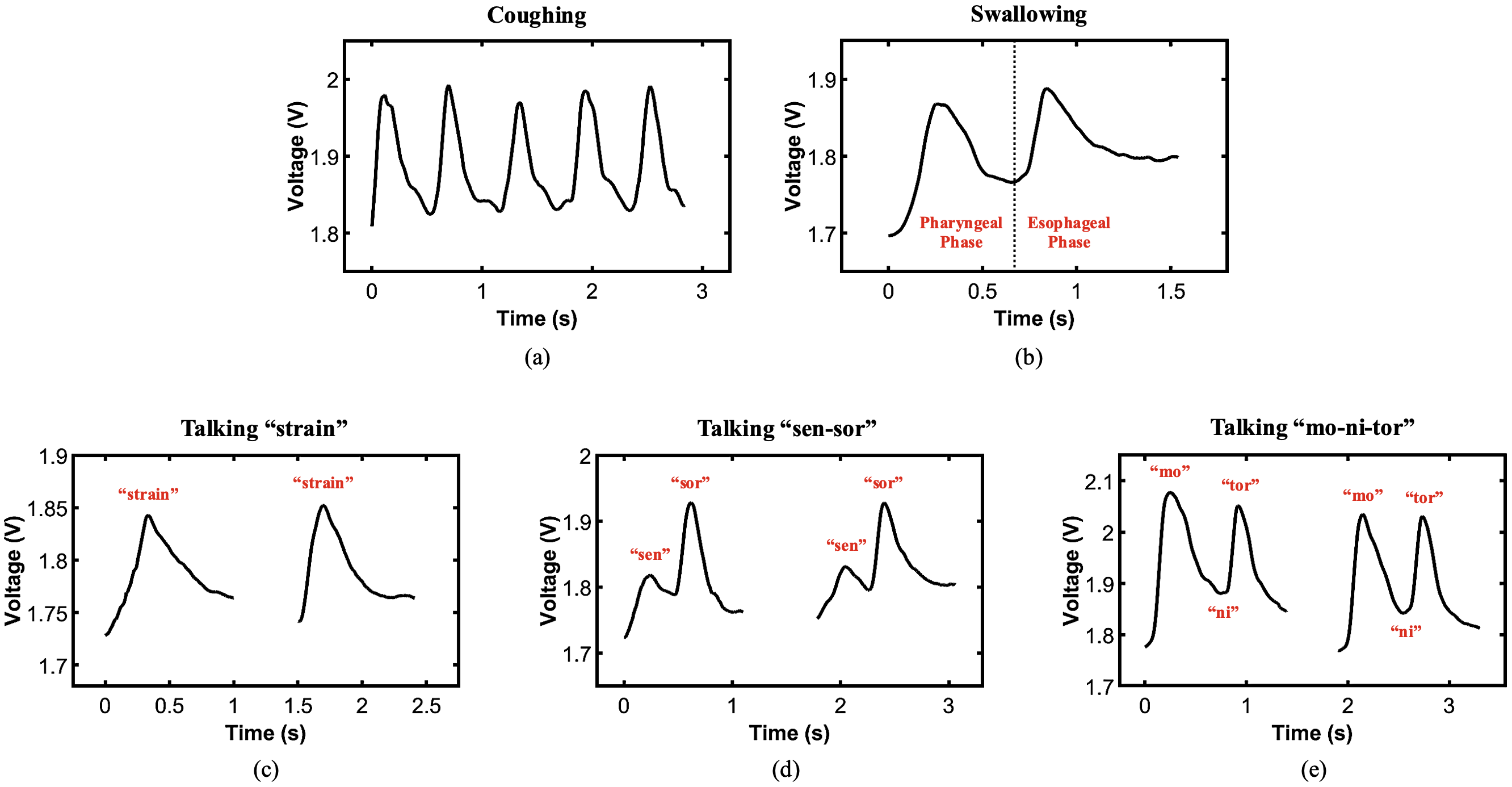}
    \caption{Neck motion measurement results. (a) Coughing. (b) Swallowing. (c) Talking the word ``strain''. (d) Talking the word ``sen-sor''. (e) Talking the word ``mo-ni-tor''.}
    \label{fig5}
\end{figure*}

Fig. \ref{fig5}a depicts the measurement results of the cough detection experiment, showing five peaks that indicate five detected cough events. Each cough event results in movement of the Adam’s apple, causing the deformation of the FPY sensor placed directly on it. The signal amplitude correlates with the intensity of the cough, which affects the level of the sensor deformation. The measurement results align with the cough signal characteristics, featuring a single peak in each cough event. Cough detection is essential for assessing the clinical condition of individuals with cough by analysing the intensity and frequency of their cough. This information is important for evaluating the severity of the cough, enabling the determination of appropriate treatment for the patient.

The following experiment relates to the swallowing detection. A normal swallowing signal has two peaks in a single swallowing event, with the second peak having slightly greater amplitude. The measurement results of the swallowing detection presented in Fig. \ref{fig5}b show a signal that corresponds to these features. The swallowing process consists of three phases: oral, pharyngeal, and esophageal. The oral phase begins when the bolus is prepared in the mouth, followed by the pharyngeal phase during which the bolus is pushed into the esophagus, and concludes in the esophageal phase when the bolus has reached the esophagus to be transported to the stomach \cite{ref-44}. The first peak of the signal results from the upward movement of the Adam’s apple during the pharyngeal phase, whereas the second peak is caused by the downward movement of the Adam’s apple to return to its initial position at the esophageal phase. The downward movement of the Adam’s apple is accompanied by a slight forward movement, causing a larger sensor deformation and producing a greater signal amplitude. The measurement results also show that the measured swallowing signal has a period of 1.5 seconds, which corresponds to the normal swallowing signal period. Swallowing detection can be used to identify swallowing difficulties (dysphagia), a common symptom of laryngeal or pharyngeal cancer. Early detection of dysphagia could reduce the risk of cancer development.

The last evaluation in the neck measurements corresponds to talking detection. Talking detection is used to identify waveforms produced from the pronunciation of a word, based on the movement of the Adam’s apple. Talking signals have characteristics where the number of signal peaks corresponds with the number of syllables pronounced. Fig. \ref{fig5}c presents the measurement results for the pronunciation of the word ``strain''. Based on the measurements conducted, both signals have a single peak, as the word ``strain'' has only one syllable. Fig. \ref{fig5}d presents the measurement results for the pronunciation of the word ``sen-sor'', showing that both signals have two peaks. The first peak that results from the pronunciation of the syllable ``sen'' has a smaller amplitude, as it generates a smaller movement of the Adam’s apple in comparison to the pronunciation of the syllable ``sor''. Then, the measurement results for the pronunciation of the word ``mo-ni-tor'' are presented in Fig. \ref{fig5}e. Although the pronunciation of three syllables, each signal has only two peaks. The pronunciation of the syllable ``ni'' does not cause the Adam’s apple to move, thereby unable to cause the deformation of the sensor. In conclusion, the signal waveform shows a clear distinction in the pronunciation of different words, while demonstrating significant overlapping in the pronunciation of the same words. These results indicate that the FPY sensor is reliable for talking detection. The integration of measurement results with pattern recognition methods enables its application in voice assistance and human-machine interaction.

\subsection{Finger Bending Measurement}
\begin{figure}[t]
    \centering
    \includegraphics[width=3.35in]{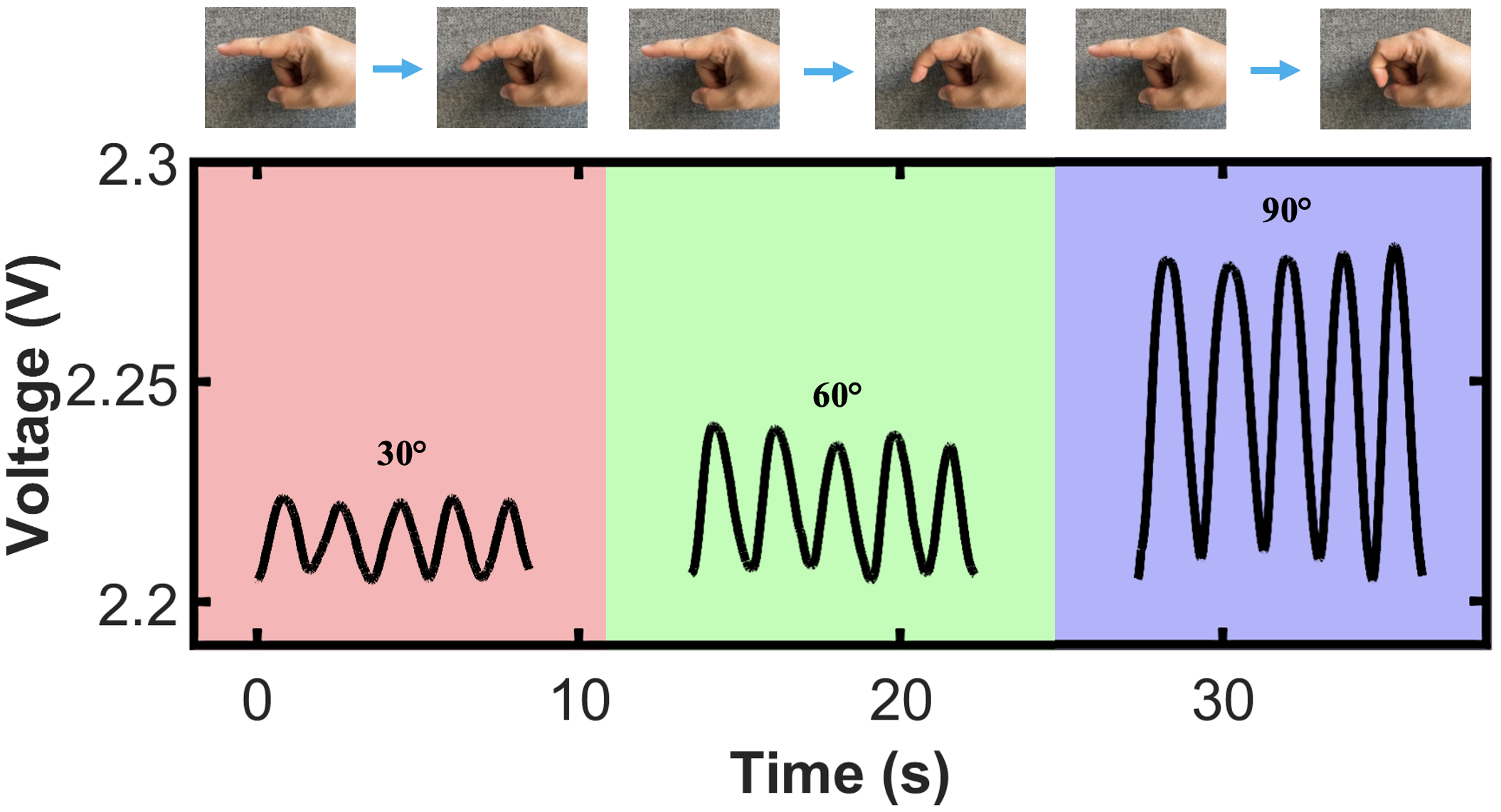}
    \caption{Finger bending measurement results at bending angles of 30°, 60°, and 90°.}
    \label{fig6}
\end{figure}

The finger bending measurement results at three different bending angles (30°, 60°, 90°) are shown in Fig. \ref{fig6}. At each examined finger joint angle, the measurement results show the peak amplitude stability during five repeated bends at the same angle. Additionally, an increase in the finger joint angle results in a higher strain level measured by the FPY sensor, thereby producing a proportionally greater signal amplitude. These results are consistent with the characteristics of finger bending signals reported in related studies. The detection of finger joint angles can be further applied in various fields, including finger gesture recognition for application in a human-machine interface for the control of a robotic arm.

\subsection{Respiratory Signal Measurement}
The respiratory process consists of the inhalation and exhalation phases. Inhalation corresponds to the intake of air into the body, whereas exhalation denotes the expulsion of air. All of these processes involve the diaphragm, a muscle located between the thoracic cavity and the abdominal cavity. During inhalation, the diaphragm contracts, resulting in the expansion of the abdomen and stretching of the FPY sensor attached to its circumference. In the exhalation, the opposite process occurs, with the abdomen returning to its initial condition \cite{ref-45}.

Fig. \ref{fig7}a depicts the measurement results of the respiratory signal during normal, deep, and rapid breathing. Based on these results, the FPY sensor effectively detects abdominal movement during respiration. The measurement results show continuous signals with a sinusoidal pattern and distinctly represent variations in amplitude and period between normal, deep, and rapid breathing. Deep breathing shows the greatest amplitude due to maximum abdominal expansion during the respiratory process, resulting in the highest degree of sensor deformation. Moreover, deep breathing has the longest signal period corresponding to the duration of the respiratory cycle. In contrast, rapid breathing has the smallest signal amplitude and the shortest period for each respiratory cycle. The inhalation and exhalation phases of the signal are shown in the single period signal presented in Fig. \ref{fig7}b. Additionally, the spectrum frequency analysis of each breathing type finds that the dominant frequencies for normal, deep, and rapid breathing are 0.3 Hz, 0.2 Hz, and 0.6 Hz, respectively. All of these measurement characteristics align with those of related studies. Continuous measurement of respiratory signals is applicable in monitoring systems for the early detection of respiratory disorders, including asthma, sleep apnea, and chronic obstructive pulmonary disease (COPD).

\subsection{Arterial Blood Pressure (ABP) Waveform Measurement}
The ABP waveform represents the pressure signal at the arterial site generated by the circulatory cycle. The circulatory cycle consists of two phases: systole and diastole. Systole is characterized by the contraction of the heart, when the left ventricle pumps blood throughout the body. Diastole denotes the phase during which the heart relaxes and fills with blood. Morphologically, the systolic phase is identified by an increase in pressure to its peak (systolic peak), followed by a decrease along with the closure of the aortic valve (dicrotic notch). The diastolic phase, starting at the dicrotic notch and concluding at the cycle’s end. This phase is recognizable by a dicrotic peak resulting from blood backflow when the aortic valve is entirely closed, followed by a decline in pressure to the endpoint \cite{ref-46}. Periodic and continuous arterial pressure during the circulatory cycle causes contact between the skin on the arterial site and the pressure mode FPY sensor attached to it, resulting in periodic changes in the sensor’s electrical resistance.

The ABP waveform measurement results presented in Fig. \ref{fig8}a show that the FPY sensor is capable of continuously detecting arterial pressure. The signal has great stability and clarity in identifying the systolic peak, dicrotic notch, and dicrotic peak, as illustrated in Fig. \ref{fig8}b. The magnitude of the arterial pressure determines the amplitude of the signal, whereas the period of each cycle depends on the duration of each circulatory cycle. Frequency spectrum analysis indicates that the dominant frequency of the measured ABP waveform is 1.16 Hz, which is within the normal heart rate frequency. The successful detection of arterial pressure, a low-magnitude signal, validates the high sensitivity of the designed sensor. The implementation of continuous non-invasive blood pressure waveform measurement is essential for the timely monitoring of cardiac health problems.

\begin{figure}[t]
    \centering
    \includegraphics[width=3.35in]{Figures/Figure7.png}
    \caption{Respiratory signal measurement results. (a) Respiratory signal of normal, deep, and rapid breathing. (b) Single period respiratory signal showing the inhalation and exhalation phases of the signal.}
    \label{fig7}
\end{figure}

\subsection{Baseline Drift Calculation}
The results of the quantitative evaluation of the physiological signal measurement are summarized in Table \ref{table1}, which contains MAE and SD of baseline drift, average and SD of signal amplitude, and the average of the percentage of baseline drift relative to signal amplitude. Based on the baseline drift results, all physiological signals have a low MAE. These results indicate a minor baseline drift occurring in each measurement. The highest MAE was identified in rapid breathing measurements, with a value of 0.0593 ± 0.0545. Among all physiological measurements, rapid breathing is the most susceptible to sensor position shifting caused by extensive abdominal movement during measurement. This condition directly affects the baseline stability, although the observed baseline drift is still acceptable.

\begin{figure}[t]
    \centering
    \includegraphics[width=2.8in]{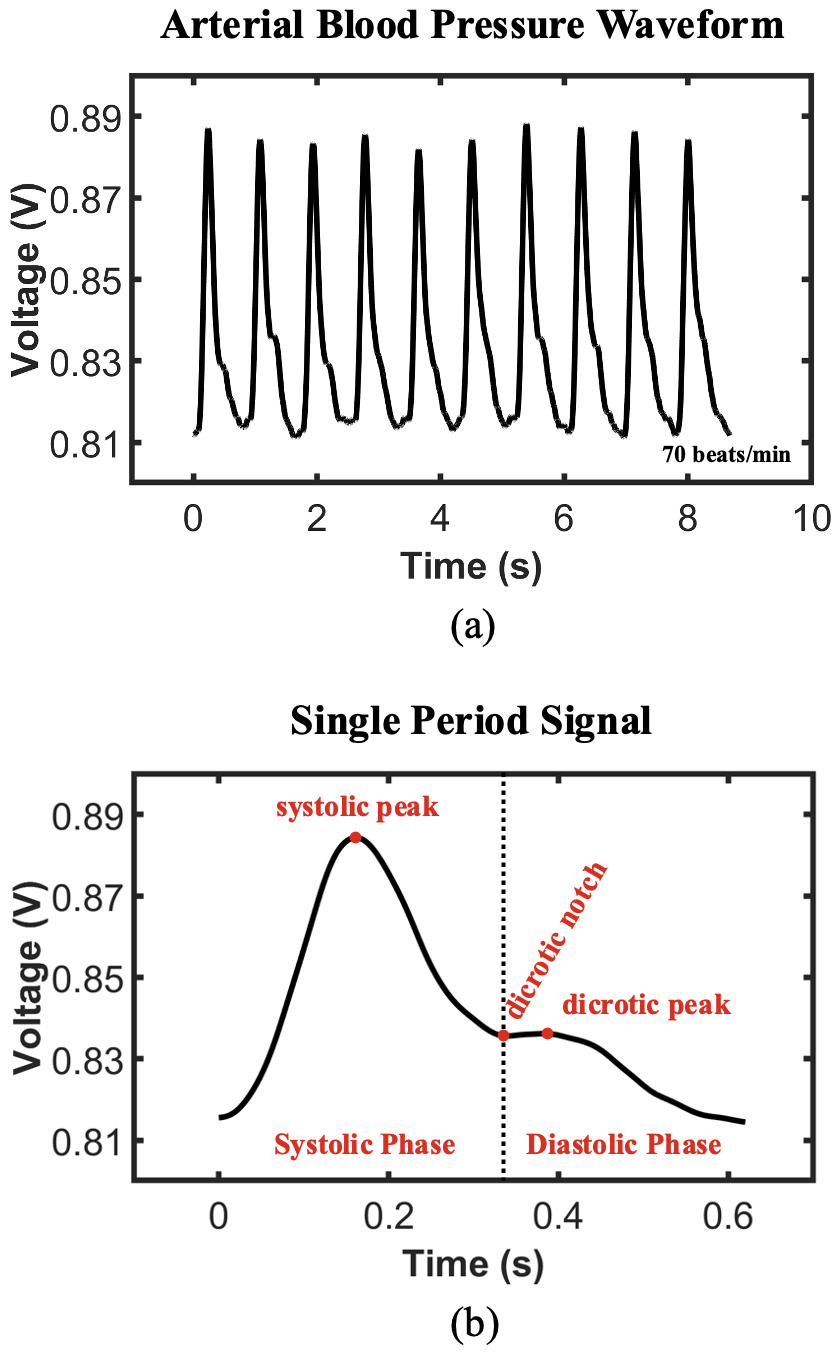}
    \caption{ABP waveform measurement results. (a) ABP waveform recorded over a duration of 10 seconds. (b) Single period ABP waveform showing the systolic and diastolic phases of the signal.}
    \label{fig8}
\end{figure}

The average amplitude of each measurement confirms the proportionality between the mechanical movement of the body and the measured signal amplitude. The highest average amplitudes for neck motion, finger bending, and respiratory signal measurement were obtained in talking the word ``mo-ni-tor'', finger bending at a 90° angle, and deep breathing, respectively. The ABP waveform, characterized by its low magnitude, had the lowest amplitude with an average of 0.1637. The low SD of each measurement indicates high repeatability of the sensor in detecting sensor deformation under the same measurement conditions.

According to the average percentage of RBD, only one measurement had percentages above 20\%, which is the rapid breathing signal (22.86\%). The overall results suggest that the FPY sensor has acceptable baseline stability. It is important to highlight that baseline drift does not significantly affect signal morphology, as shown in the qualitative evaluation of signal characteristics, especially when the baseline drift is minor, as observed in this study. For in-depth physiological signal analysis, baseline drift can be corrected using specific algorithms.

\begin{table}
\caption{\textbf{Baseline Drift (MAE \& SD), Signal Amplitude (Average \& SD), and the Percentage Average of RBD for Each Physiological Signal Measurement.}}
\label{table1}
\setlength{\tabcolsep}{4pt}
\centering
\renewcommand{\arraystretch}{1.2}
\begin{tabular}{cccccc}
\Xhline{2\arrayrulewidth}
\multirow{2}{*}{\textbf{Measurements}} & \multicolumn{2}{c}{\textbf{Baseline Drift}} & \multicolumn{2}{c}{\textbf{Signal Amplitude}} & \multirow{2}{*}{\textbf{RBD (\%)}} \\
\ & \textbf{MAE} & \textbf{SD} & \textbf{Average} & \textbf{SD} & \\
\hline
\multicolumn{1}{l}{\textbf{Neck Motion}} & & & & & \\
\multicolumn{1}{l}{Coughing} & 0.0225 & 0.0144 & 0.3567 & 0.0341 & 6.35 \\
\multicolumn{1}{l}{Swallowing} & 0.0242 & 0.0209 & 0.1646 & 0.0332 & 16.73 \\
\multicolumn{1}{l}{Talking ``strain''} & 0.0300 & 0.0093 & 0.2433 & 0.0132 & 11.25 \\
\multicolumn{1}{l}{Talking ``sen-sor''} & 0.0293 & 0.0093 & 0.3140 & 0.0311 & 7.87 \\
\multicolumn{1}{l}{Talking ``mo-ni-tor''} & 0.0406 & 0.0164 & 0.5508 & 0.0318 & 7.17 \\
\hline
\multicolumn{1}{l}{\textbf{Finger Bending}} & & & & & \\
\multicolumn{1}{l}{$30^\circ$} & 0.0075 & 0.0063 & 0.1690 & 0.0358 & 5.00 \\
\multicolumn{1}{l}{$60^\circ$} & 0.0085 & 0.0059 & 0.3273 & 0.0258 & 2.55 \\
\multicolumn{1}{l}{$90^\circ$} & 0.0110 & 0.0087 & 0.7665 & 0.0256 & 1.46 \\
\hline
\multicolumn{1}{l}{\textbf{Respiratory Signal}} & & & & & \\
\multicolumn{1}{l}{Normal} & 0.0317 & 0.0143 & 0.5145 & 0.0473 & 6.41 \\
\multicolumn{1}{l}{Deep} & 0.0248 & 0.0114 & 0.6211 & 0.0627 & 3.87 \\
\multicolumn{1}{l}{Rapid} & 0.0593 & 0.0545 & 0.2427 & 0.0557 & 22.86 \\
\hline
\multicolumn{1}{l}{\textbf{ABP Waveform}} & & & & & \\
\multicolumn{1}{l}{ABP Waveform} & 0.0051 & 0.0029 & 0.1637 & 0.0338 & 3.30 \\
\Xhline{2\arrayrulewidth}
\end{tabular}
\end{table}

\section{Conclusions}
\label{sec:conclusions}
In this study, we developed and optimized the design of an FPY sensor based on conductive TPU filament for multiphysiological signal measurement within the body. The optimal sensor design is achieved through the implementation of FPY as a bonding pattern on two parallel polyester yarns, utilizing a tight pattern and minimal length FPY while sufficiently covering the measurement point. The prototype sensor has the ability to operate in two measurement modes: strain and pressure. The strain mode is capable of measuring neck motion, finger bending, and the respiratory signal. Meanwhile, the pressure mode is able to detect low-magnitude pressure in ABP waveform measurement. The measurement results of each physiological signal show that the morphological characteristics of the measured signals are identical to the characteristics defined in related studies. The reliability of the FPY sensor in accurately detecting all measured physiological signals demonstrates its wide working range and high sensitivity. Acceptable measurement stability can also be identified by the minor baseline drift of each measurement, as well as its small percentage relative to its amplitude. Furthermore, the high repeatability of the sensor is shown by the low standard deviation of the signal amplitude, which further illustrates proportionality to the detected strain or pressure intensity. We believe that the features of the FPY sensor we have developed could serve as an innovative approach in the physiological signal monitoring for various applications, such as personalized healthcare and sports.

\ifCLASSOPTIONcaptionsoff
  \newpage
\fi

\begin{IEEEbiographynophoto}{Rizal Maulana} received the M.S. degree in electrical engineering from the National Central University, Taoyuan, Taiwan, and Brawijaya University, Malang, Indonesia (Double Degree Program), in 2014. He is currently pursuing the Ph.D. degree with the Doctoral School of Sciences and Technology, Pázmány Péter Catholic University, Budapest, Hungary. He began his career as a Lecturer with Brawijaya University, Malang, Indonesia, in 2015. His research interests include biomedical engineering and robotics.
\end{IEEEbiographynophoto}

\begin{IEEEbiographynophoto}{Ádám Rák} received the Ph.D. degree from the Multidisciplinary Doctoral School of Engineering and Natural Sciences, Pázmány Péter Catholic University, Budapest, Hungary, in 2014. Since 2022, he has been a Researcher at the Faculty of Information Technology and Bionics, Pázmány Péter Catholic University, Budapest, Hungary. His research interests include parallel and GPU computing, FPGA design, cellular neural networks, machine learning, artificial intelligence, and numerical quantum chemistry.
\end{IEEEbiographynophoto}

\begin{IEEEbiographynophoto}{Sándor Földi} received the Ph.D. degree in information science from the Faculty of Information Technology and Bionics at Pázmány Péter Catholic University, Budapest, Hungary, in 2020. He is currently a senior lecturer at the Faculty of Information Technology and Bionics, Pázmány Péter Catholic University, Budapest, Hungary. His research interests include blood pressure monitoring, pulse diagnosis, biomedical signal processing, sensor technologies, rehabilitation robotics, and biomechanics.
\end{IEEEbiographynophoto}

\begin{IEEEbiographynophoto}{György Cserey} (Member, IEEE) received the Ph.D. degree in infobionics from Pázmány Péter Catholic University, Budapest, Hungary, in 2006. In 2003 he spent one year as a Visiting Researcher at the University of Notre Dame, Notre Dame, Indiana, United States. He is currently a Professor and the Dean at the Faculty of Information Technology and Bionics, Pázmány Péter Catholic University, Budapest, Hungary. His publication statistics include more than 100 papers and nine patents. His research interests include artificial intelligence, sensory robotics, parallel computing and neuro-bio-inspired systems.
\end{IEEEbiographynophoto}

\begin{thebibliography}{00}
\bibitem{ref-1} C. Xu, J. Chen, Z. Zhu, M. Liu, R. Lan, X. Chen, W. Tang, Y. Zhang, and H. Li, ``Flexible pressure sensors in human-machine interface applications,'' \emph{Small}, vol. 20, no. 15, Apr. 2024.

\bibitem{ref-2} W. Xiao, X. Cai, A. Jadoon, Y. Zhou, Q. Gou, J. Tang, X. Ma, W. Wang, and J. Cai, ``High-performance graphene flexible sensors for pulse monitoring and human-machine interaction,'' \emph{ACS Appl. Mater. Interfaces}, vol. 16, no. 25, pp. 32445–32455, Jun. 2024.

\bibitem{ref-3} M. Zhong, L. Zhang, X. Liu, Y. Zhou, M. Zhang, Y. Wang, L. Yang, and D. Wei, ``Wide linear range and highly sensitive flexible pressure sensor based on multistage sensing process for health monitoring and human-machine interfaces,'' \emph{Chem. Eng. J.}, vol. 412, May 2021.

\bibitem{ref-4} G. Yao, C. Yin, Q. Wang, T. Zhang, S. Chen, C. Lu, K. Zhao, W. Xu, T. Pan, M. Gao, and Y. Lin, ``Flexible bioelectronics for physiological signals sensing and disease treatment,'' \emph{J. Materiomics}, vol. 6, no. 2, pp. 397-413, Jun. 2020.

\bibitem{ref-5} T. Yang, W. Deng, X. Chu, X. Wang, Y. Hu, X. Fan, J. Song, Y. Gao, B. Zhang, G. Tian, D. Xiong, S. Zhong, L. Tang, Y. Hu, and W. Yang, ``Hierarchically microstructure-bioinspired flexible piezoresistive bioelectronics,'' \emph{ACS Nano}, vol. 15, no. 7, pp. 11555–11563, Jul. 2021.

\bibitem{ref-6} C. Chitrakar, E. Hedrick, L. Adegoke, and M. Ecker, ``Flexible and stretchable bioelectronics,'' \emph{Materials}, vol. 15, no. 5, p.1664, Feb. 2022.

\bibitem{ref-7} S. Zhang, A. Chhetry, M.A. Zahed, S. Sharma, C. Park, S. Yoon, and J.Y. Park, ``On-skin ultrathin and stretchable multifunctional sensor for smart healthcare wearables,'' \emph{npj Flex. Electron.}, vol. 6, no. 1, p. 11, Feb. 2022.

\bibitem{ref-8} A. Yu, M. Zhu, C. Chen, Y. Li, H. Cui, S. Liu, and Q. Zhao, ``Implantable flexible sensors for health monitoring,'' \emph{Adv. Healthcare Mater.}, vol. 13, no. 2, Jan. 2024.

\bibitem{ref-9} D.N. Elsheakh, O.M. Fahmy, M. Farouk, K. Ezzat, and A.R. Eldamak, ``An early breast cancer detection by using wearable flexible sensors and artificial intelligent,'' \emph{IEEE Access}, vol. 12, pp. 48511-48529, Apr. 2024.

\bibitem{ref-10} Y. Wang, H. Haick, S. Guo, C. Wang, S. Lee, T. Yokota, and T. Someya, ``Skin bioelectronics towards long-term, continuous health monitoring,'' \emph{Chem. Soc. Rev.}, vol. 51, no. 9, pp. 3759-3793, May 2022.

\bibitem{ref-11} K. Meng, X. Xiao, W. Wei, G. Chen, A. Nashalian, S. Shen, X. Xiao, and J. Chen, ``Wearable pressure sensors for pulse wave monitoring,'' \emph{Adv. Mater.}, vol. 34, no. 21, May 2022.

\bibitem{ref-12} U.P. Claver and G. Zhao, ``Recent progress in flexible pressure sensors based electronic skin,'' \emph{Adv. Eng. Mater.}, vol. 23, no. 5, p. 2001187, May 2021.

\bibitem{ref-13} Z. Shen, F. Liu, S. Huang, H. Wang, C. Yang, T. Hang, J. Tao, W. Xia, and X. Xie, ``Progress of flexible strain sensors for physiological signal monitoring,'' \emph{Biosens. Bioelectron.}, vol. 211, Sep. 2022.

\bibitem{ref-14} M.A. Signore, G. Rescio, L. Francioso, F. Casino, and A. Leone, ``Aluminum nitride thin film piezoelectric pressure sensor for respiratory rate detection,'' \emph{Sensors}, vol. 24, no. 7, p. 2071, Mar. 2024.

\bibitem{ref-15} Y.G. Kim, J.H. Song, S. Hong, and S.H. Ahn, ``Piezoelectric strain sensor with high sensitivity and high stretchability based on kirigami design cutting,'' \emph{npj Flex. Electron.}, vol. 6, Jun. 2022.

\bibitem{ref-16} P. Yang, Y. Shi, S. Li, X. Tao, Z. Liu, X. Wang, Z.L. Wang, and X. Chen, ``Monitoring the degree of comfort of shoes in-motion using triboelectric pressure sensors with an ultrawide detection range,'' \emph{ACS Nano}, vol. 16, no. 3, pp. 4654-4665, Mar. 2022.

\bibitem{ref-17} G. Li, S. Wu, Z. Sha, L. Zhao, D. Chu, C.H. Wang, and S. Peng, ``A triboelectric nanogenerator powered piezoresistive strain sensing technique insensitive to output variations,'' \emph{Nano Energy}, vol. 108, p. 108185, Apr. 2023.

\bibitem{ref-18} M.Q. Mehmood, M.H. Zulfiqar, A.K. Goyal, M.S. Malik, W.T. Khan, M.A. Khan, M. Zubair, and Y. Massoud, ``Lithography-based fabricated capacitive pressure sensitive touch sensors for multimode intelligent HMIs,'' \emph{IEEE Access}, vol. 11, pp. 127411-127421, Nov. 2023.

\bibitem{ref-19} H. Xu, Y. Lv, D. Qiu, Y. Zhou, H. Zeng, and Y. Chu, ``An ultra-stretchable, highly sensitive and biocompatible capacitive strain sensor from an ionic nanocomposite for on-skin monitoring,'' \emph{Nanoscale}, vol. 11, no. 4, pp. 1570-1578, Jan. 2019.

\bibitem{ref-20} M. Cao, J. Su, S. Fan, H. Qiu, D. Su, and L. Li, ``Wearable piezoresistive pressure sensors based on 3d graphene,'' \emph{Chem. Eng. J.}, vol. 406, p. 126777, Feb. 2021.

\bibitem{ref-21} Y. Yi, B. Wang, X. Liu, and C. Li, ``Flexible piezoresistive strain sensor based on CNTs–polymer composites: a brief review,'' \emph{Carbon Lett.}, vol. 32, pp. 713-726, Feb. 2022.

\bibitem{ref-22} W. Zhong, W. Liu, Y. Ke, K. Jia, X. Ming, M. Li, D. Wang, Y. Chen, and H. Jiang, ``Knotted fiber-based strain sensors with tunable sensitivity and a sensing region for monitoring wearable physiological signals and human motion,'' \emph{J. Mater. Chem. C}, vol. 11, no. 42, pp. 14796-14804, Oct. 2023.

\bibitem{ref-23} S. Han, S. Li, X. Fu, S. Han, H. Chen, L. Zhang, J. Wang, and G. Sun, ``Research progress of flexible piezoresistive sensors based on polymer porous materials,'' \emph{ACS Sens.}, vol. 9, no. 8, pp. 3848-3863, Aug. 2024.

\bibitem{ref-24} Y. Zhu, S. Jiang, Y. Xiao, J. Yu, L. Sun, and W. Zhang, ``A flexible three-dimensional force sensor based on PI piezoresistive film,'' \emph{J. Mater. Sci: Mater. Electron.}, vol. 29, pp. 19830-19839, Sep. 2018.

\bibitem{ref-25} V. Pandey, A. Mandal, S. Sisle, M.P. Gururajan, and R.O. Dusane, ``Piezoresistive pressure sensor using nanocrystalline silicon thin film on flexible substrate,'' \emph{Sens. Actuators A: Phys.}, vol. 316, p. 112372, Dec. 2020.

\bibitem{ref-26} N.A. Choudhry, A. Rasheed, S. Ahmad, L. Arnold, and L. Wang, ``Design, development and characterization of textile stitch-based piezoresistive sensors for wearable monitoring,'' \emph{IEEE Sens. J.}, vol. 20, no. 18, pp. 10485-10494, Sep. 2020.

\bibitem{ref-27} Y. Yang, Y. Liu, and R. Yin, ``Fiber/yarn and textile-based piezoresistive pressure sensors,'' \emph{Adv. Fiber Mater.}, vol. 7, pp. 34-71, Feb. 2025.

\bibitem{ref-28} M.M. Hosen, A. Ferdous, and S. Ahmed, ``Easily scalable and highly flexible machine knitted resistive pressure sensor for smart textile applications,'' \emph{J. Ind. Text.}, vol. 54, Aug. 2024.

\bibitem{ref-29} A. Sadeqi, H.R. Nejad, F. Alaimo, H. Yun, M. Punjiya, and S.R. Sonkusale, ``Washable smart threads for strain sensing fabrics,'' \emph{IEEE Sens. J.}, vol. 18, no. 22, pp. 9137-9144, Nov. 2018.

\bibitem{ref-30} Y. Ke, K. Jia, W. Zhong, X. Ming, H. Jiang, J. Chen, X. Ding, M. Li, Z. Lu, and D. Wang, ``Wide-range sensitive all-textile piezoresistive sensors assembled with biomimetic core-shell yarn via facile embroidery integration,'' \emph{Chem. Eng. J.}, vol. 435, p. 135003, May 2022.

\bibitem{ref-31} Q. Zhang, L. Shen, P. Liu, P. Xia, J. Li, H. Feng, C. Liu, K. Xing, A. Song, M. Li, X. Yang, and Y. Huang, ``Highly sensitive resistance-type flexible pressure sensor for cuffless blood-pressure monitoring by using neural network techniques,'' \emph{Compos. B: Eng.}, vol. 226, Dec. 2021.

\bibitem{ref-32} S. Földi, T. Horváth, F. Zieger, P. Sótonyi, and G. Cserey, ``A novel non‑invasive blood pressure waveform measuring system compared to millar applanation tonometry,'' \emph{J. Clin. Monit. Comput.}, vol. 32, pp. 717-727, Aug. 2018.

\bibitem{ref-33} J. Tolvanen, J. Hannu, and H. Jantunen, ``Stretchable and washable strain sensor based on cracking structure for human motion monitoring,'' \emph{Sci. Rep.}, vol. 8, Sep. 2018.

\bibitem{ref-34} T. Otoshi, T. Nagano, S. Izumi, D. Hazama, N. Katsurada, M. Yamamoto, M. Tachihara, K. Kobayashi, and Y. Nishimura, ``A novel automatic cough frequency monitoring system combining a triaxial accelerometer and a stretchable strain sensor,'' \emph{Sci. Rep.}, vol. 11, p. 9973, May 2021.

\bibitem{ref-35} J. Ramírez, D. Rodriquez, F. Qiao, J. Warchall, J. Rye, E. Aklile, A.S.C. Chiang, B.C. Marin, P.P. Mercier, C.K. Cheng, K.A. Hutcheson, E.H. Shinn, and D.J. Lipomi, ``Metallic nanoislands on graphene for monitoring swallowing activity in head and neck cancer patients,'' \emph{ACS Nano}, vol. 12, no. 6, pp. 5913-5922, Jun. 2018.

\bibitem{ref-36} B. Polat, T. Rafeedi, L. Becerra, A.X. Chen, K. Chiang, V. Kaipu, R. Blau, P.P. Mercier, C.K. Cheng, and D.J. Lipomi, ``External measurement of swallowed volume during exercise enabled by stretchable derivatives of PEDOT:PSS, graphene, metallic nanoparticles, and machine learning,'' \emph{Adv. Sensor Res.}, vol. 2, no. 4, p. 2200060, Apr. 2023.

\bibitem{ref-37} X. Sun, Q. Wang, J. Zhan, T. Yang, Y. Zhao, C.K. Sun, M. Aisha, M. Guo, S. Tang, H. Zhao, L. Wang, and J. Liu, ``Superhydrophobic conductive suede fabrics based on carboxylated multiwalled carbon nanotubes and polydopamine for wearable pressure sensors,'' \emph{ACS Appl. Nano Mater.}, vol. 6, no. 12, pp. 10746–10757, Jun. 2023.

\bibitem{ref-38} J. Wang, S. Ren, X. Jia, and Y. Jia, ``Liquid metal and carbon nanofiber-based strain sensor for monitoring gesture, voice, and physiological signals,'' \emph{ACS Appl. Nano Mater.}, vol. 7, no. 2, pp. 1664–1673, Jan. 2024.

\bibitem{ref-39} A. Wang, M. Hu, L. Zhou, and X. Qiang, ``Self-powered wearable pressure sensors with enhanced piezoelectric properties of aligned P(VDF-TrFE)/MWCNT composites for monitoring human physiological and muscle motion signs,'' \emph{Nanomaterials}, vol. 8, no. 12, p. 1021, Dec. 2018.

\bibitem{ref-40} Z. Pang, Y. Zhao, N. Luo, D. Chen, and M. Chen, ``Flexible pressure and temperature dual-mode sensor based on buckling carbon nanofibers for respiration pattern recognition,'' \emph{Sci. Rep.}, vol. 12, p. 17434, Oct. 2022.

\bibitem{ref-41} P.R. Priya, S.K. Dubey, and S. Goel, ``Noninvasive clean room free printed piezoresistive breath sensor for point of care application,'' \emph{IEEE Sens. J.}, vol. 23, no. 12, pp. 13621-13628, Jun. 2023.

\bibitem{ref-42} A. Mandal, A. Morali, M. Skorobogatiy, and S. Bodkhe, ``3D printing of polyvinylidene fluoride-based piezoelectric sensors for noninvasive continuous blood pressure monitoring,'' \emph{Adv. Eng. Mater.}, vol. 26, no. 1, p. 2301292, Jan. 2024.

\bibitem{ref-43} N. Wu, S. Chen, S. Lin, W. Li, Z. Xu, F. Yuan, L. Huang, B. Hu, and J. Zhou, ``Theoretical study and structural optimization of a flexible piezoelectret-based pressure sensor,'' \emph{J. Mater. Chem. A.}, vol. 6, no. 12, pp. 5065-5070, Feb. 2018.

\bibitem{ref-44} J. Walton and P. Silva, ``Physiology of swallowing,'' \emph{Surgery (Oxf).}, vol. 42, no. 9, pp. 638-643, Sep. 2024.

\bibitem{ref-45} H. Hamasaki, ``Effects of diaphragmatic breathing on health: a narrative review,'' \emph{Medicines}, vol. 7, no. 10, p. 65, Oct. 2020.

\bibitem{ref-46} Q. Tang, Z. Chen, R. Ward, and M. Elgendi, ``Synthetic photoplethysmogram generation using two gaussian functions,'' \emph{Sci. Rep.}, vol. 10, p. 13883, Aug. 2020.
\end{thebibliography}
\end{document}